# Mechanical Properties and Failure Behavior of Phosphorene with Grain Boundaries

V. Sorkin[*] and Y.W. Zhang[†]

Institute of High Performance Computing, A*STAR, Singapore 138632

## Abstract

Using density functional tight-binding method, we studied the effect of grain boundaries on the mechanical properties and failure behavior of phosphorene. We found that the large angle tilt boundaries with a higher density of (5|7) defect pairs (oriented along the AC direction) are stronger than the low-angle tilt boundaries with a lower defect density, and similarly the large angle boundaries with a higher density of (4|8) defect pairs (oriented along the ZZ direction) are stronger than the low-angle boundaries with a lower defect density. The failure is due to the rupture of the most pre-strained bonds in the heptagons of the (5|7) defect pair or octagons of the (4|8) pairs. The large-angle grain boundaries are better off in accommodating the pre-strained bonds in heptagons and octagons defects, leading to a higher failure stress and strain. The results cannot be described by Griffith-type fracture mechanics criterion since it does not take into account the bond pre-stretching. Interestingly, these anomalous mechanical and failure characteristics of tilt grain boundaries in phosphorene are also shared by graphene and hexagonal born-nitride, signifying that they may be universal for 2D materials. The findings revealed here may be useful in tuning the mechanical properties of phosphorene via defect engineering for specific applications.

## 1. Introduction

As a new class of materials, 2D materials have received much attention due to their fascinating electronic, thermal and mechanical properties, which can be used in a variety of promising novel applications in nano-electronics, flexible electronics and energy conversion. Among this fast growing family of 2D materials, phosphorene [1,2] takes a special place as it bridges the gap between semi-metallic graphene [3], insulating boron-nitride [4] and transition metal dichalcogenides [5]. Since phosphorene possesses high direction-dependent carrier mobilities and anisotropy in thermal and electronic properties, it is most suitable for applications in nano-electronics as field-effect transistors and as a building block in logic circuits [6], and also in gas sensing, thermoelectrics, Li-ion batteries, and solar-cell devices [7–9]. Furthermore, its sensitivity to applied strain[10] also makes it a particularly attractive material for flexible electronics.

---

[*] Email address: sorkinv@ihpc.a-star.edu.sg
[†] Email address: zhangyw@ihpc.a-star.edu.sg

While the mechanical properties of pristine phosphorene [11], such as its large non-linear deformation and failure behavior under tensile strain, have been studied extensively [12], the effects of defects on the mechanical properties of phosphorene have just stated to attract attention. Defects are intrinsically present in 2D materials and they can also be intentionally introduced to modify their mechanical properties. In general, there are two types of intrinsic defects in 2D materials: point defects (vacancies, interstitials, dislocations and topological defects) and line defects (grain boundaries). Grain boundaries exemplify a class of defects, which are defined by a structural topological invariant that does not alter upon local modifications of the lattice [13]. Due to the difference in lattice structures and bonding energies, the configurations of these defects may take different forms in a variety of 2D materials. The effects of grain boundaries (GBs) on other 2D materials, especially graphene [14,15], boron-nitride (BN) [16,17] and transition-metal-dichalcogenides (TMDs) [5,18] have been extensively explored.

Investigation on the effects of GBs on the mechanical properties of graphene ended with a few surprising discoveries [19–21]. It was found that its failure always starts from the bond shared by hexagon–heptagon rings. Moreover, GBs with large tilt angles (formed by closely spaced dislocations) were found to be almost as strong as pristine graphene, while small tilt angle GBs, which are characterized by larger distances between dislocations, are considerably weaker [22,23]. Recently, Wei et al. [24] demonstrated that it is not only the density of defects, but also their detailed arrangement that determines the strength enhancement and weakening in graphene. It was also found that triple junctions of graphene GBs are the nucleation centers for cracks [25]. Cracks in graphene propagate along GBs and inside grains [26], leading to failure. In general, the failure of graphene is brittle, since dislocations are completely immobile at normal conditions [21].

In contrast with the amount of investigations completed in studying the effects of defects on other 2D materials, the phosphorene is relatively less explored. Liu et al. [27] investigated energetics and electronic properties of GBs in phosphorene. They found that grain boundaries in phosphorene are electronically inactive due to the homo-elemental bonding, in contrast to hetero-elemental bonding in TMD GBs, which typically deteriorates the performance of opto-electronic devices [28]. Guo et al. [29] examined the atomic structure, thermodynamic stability, and electronic properties of phosphorene GBs composed of (5|7), (5|6|7), (5|8|7) and (5|8|8|7) rings. They confirmed that GBs do not severely affect the electronic properties of phosphorene: the band gap is preserved and the electron mobility is only slightly reduced. Jang et al. [30] investigated the temperature-dependent energetics and electronic structure of phosphorene with various GBs using density-functional theory. They identified new low-energy GBs exhibiting a range of electronic structures (from metallic to semiconducting). Using first principle calculations, Zhu et al. [31] investigated the effects of substitutional dopants by C and O on the energetics and electronic properties of phosphorene GBs. They found that GB region is reactive, and it is energetically more favorable to incorporate the dopants into the GB region than into the bulk. They also showed that the electronic and magnetic properties can be effectively tuned by the dopant atoms embedded into the GB region.

In view of the fact that grain boundaries are a fundamental defect type, it is important to understand their effects on the mechanical properties and failure behavior of phosphorene. Equally, structure integrity and failure behavior of phosphorene is one of the major concerns for any device applications. Therefore, in the present work, our main objective is to examine the grain boundary structures and their effects on the failure mechanism of phosphorene. To do so, we first constructed two types of GBs: one formed by an array of (5|7) defect pairs (oriented along the armchair direction) and another one formed by an array of (4|8) defect pairs (oriented along the zigzag direction); and then subjected them to uniaxial tensile strain. We would like to find the answers for the following questions: What is the equilibrium structure of the constructed grain boundaries? How grain boundary energy depends on the tilt angle? What is the fracture mechanism of phosphorene with linear defects? Does the tilt angle profoundly affect the deformation and failure? To answer these questions, we carried out density functional tight binding calculations.

## 2. Computational Model

We used tight-binding (TB) technique [32] to study the structure, large deformation and failure of phosphorene with grain boundaries under uniform uniaxial tensile strain. The TB technique has a very special place among the computational methods applied for nanoscale modelling of materials. On one hand, the density function theory (DFT) method, which has been extensively used to study phosphorene, is very precise, but computationally demanding. As a result, the DFT calculations are not feasible for large or even intermediate scale systems. On the other hand, the possibility to apply molecular dynamics (MD) simulations is rather restricted since the reliable and commonly accepted interatomic potential for phosphorene is not available. Consequently, empirical tight-binding technique, placed between DFT and MD in terms of computational cost and accuracy, is an appropriate method to deal with the size problem.

In our simulations of phosphorene with grain boundaries, we applied density functional tight-binding based (DFTB) method [33], which properly combines the DFT-like accuracy with the computational efficiency of TB. The DFTB is derived from DFT but uses empirical approximations to increase the computational efficiency, while preserving the accuracy [33]. The substitution of the many-body Hamiltonian of DFT with a parameterized Hamiltonian matrix is the vital approximation in the TB method [33]. In the DFTB approach, wave functions, represented by linear combinations of Slater-Koster orbitals [34], are utilized to calculate the Hamiltonian matrix elements and to model the electron density [33]. However, these matrix elements do not entirely describe the total energy of the system [35]. The remaining part is added as short-range repulsive terms represented by pair-wise potentials, obtained by fitting to DFT and experimental data [33]. Besides the short-range repulsive terms, the Kohn-Sham energy also contains the dispersion interactions (van der Waals forces) and Coulomb interactions [33]. The Coulomb interaction term describes long-range electrostatic interactions between two point charges and self-interaction contributions of a given atom (if the charges are located at the same atom) [36]. Furthermore, self-consistent charge (SCC) calculations are implemented in DFTB to improve the description of atom bonding [36]. Due to the SCC extension, the DFTB can be effectively applied to problems where deficiencies in the standard (non-SCC) TB technique are apparent [33]. Combining almost quantum mechanical accuracy and computational efficiency similar to MD methods, the DFTB comes up with remarkable opportunities to explore nano-systems inclosing a several hundreds of atoms. For example, structural, mechanical and electronic properties of phosphorene monolayer, nanoribbons and nanotubes were investigated by using DFTB [11,37,38]. In our simulations we used DFTB+, which is a fast, versatile and efficient open-source quantum mechanical simulation package [36,39].

The construction of the phosphorene samples with grain boundaries, composed of vertically oriented linear arrays of evenly spaced dislocations, is illustrated in Figure 1. Following our previous approach [11,12], we optimized the unit cell of phosphorene obtained by DFT method [9]. The optimized cell was used to construct a perfect bulk sample of phosphorene. Subsequently, we removed a group of selected atoms, forming 'wedge'-like region, from the bulk sample (see

Figure 1(a)). The elimination of these atoms is required for construction of linear arrays of dislocations. When we remove atoms along the zig-zag (ZZ) direction, the quadrilateral-octagon (4|8) defect pairs (dislocations) are introduced. When we cut atomic rows along the arm-chair (AC) direction, the pentagon-heptagon (5|7) defect pairs (dislocations) are constructed [27]. Note that graphene grain boundaries are composed of (5|7) defect pairs only since their energy is substantially lower than that of (4|8) pairs [40]. In hexagonal boron-nitride (h-BN) and transition-metal-dichalcogenides (TMDs: $MoS_2$, $WS_2$, etc.), the energy of dislocation is determined not only by topological strain, but also by the type of homo- or hetero-elemental inter-atomic bonding. The later one is substantially stronger, making (4|8) defect pairs equally preferable as (5|7) ones [27]. The existence of both (5|7) and (4|8) defect pairs in mono-elemental phosphorene is due to its puckered structure [27].

Next, the two crystalline domains of phosphorene are pivoted around a selected origin in the clock-wise and anti-clock-wise directions, respectively (see Figure 1(b)), and welded seamlessly together (with a minor bond length adjustment). The constructed defect pairs form a vertically oriented linear array of edge dislocations, which is the preferred alignment of identical edge dislocations [14]. In order to implement periodic boundary conditions in the direction perpendicular to the created grain boundary, the constructed sample is duplicated (see the highlighted atoms in Figure 1(c)). The duplicate is rotated by 180° and joined to the original one as shown in Figure 1(c). Hence, the periodic boundary conditions are implemented by constructing a sample with two grain boundaries, as generally realized in grain boundary simulations [14]. The two-dimensional edge dislocations at the grain boundaries are oriented in the opposite directions (see Figure 1(f)). In order to implement the periodic boundary conditions in the direction parallel to the grain boundaries, we need to introduce at least two (4|8) or (5|7) defects pairs. In general, an even number of defects pairs must be used in creating the grain boundaries in a phosphorene sample compatible with periodic boundary conditions. This is due to the non-planar puckered structure of phosphorene. As can be seen in Figure 1(b, side view), the phosphorene atoms are located in two (upper and lower) planes. For example, the two highlighted atoms ($A_1$, $A_4$) of the 4-member ring of (4|8) defect pair in Figure 1(d) are located in the upper plane, while the other two atoms ($A_2$, $A_3$) are in the lower plane. In the nearest neighbor (4|8) defect pair (see Figure 1(d)), the two corresponding atoms ($B_1$, $B_4$) of the 4-member ring are located in the lower plane, while the other two highlighted atoms ($B_2$, $B_3$) are located in the upper plane. In the next nearest neighbor (4|8) defect pair, the location of the atoms within the two planes is reverted to the original one, namely, the two highlighted atoms ($C_1$, $C_4$) are in the upper plane, while the other two ($C_2$, $C_3$) are in the lower plane. Thus at least two consecutive defect pairs must be included in the sample. The same alteration in the atomic positions within the two planes is evident for the (5|7) defect pairs in Figure 1(e). The highlighted atoms ($A_1$, $A_2$, $A_3$) of the (5|7) defect pair reside in the upper plane, while the highlighted atoms ($B_1$, $B_2$, $B_3$) of the nearest neighboring (5|7) defect pair are in the lower plane, meanwhile the highlighted atoms ($C_1$, $C_2$, $C_3$) of the next nearest (5|7) neighbor belong to the upper plane again. Finally, using periodic boundary conditions, we move all the sample atoms inside the computational box (see Figure 1(f)). The construction of the samples with grain boundaries can be most conveniently implemented with Materials Visualizer tools of "Materials Studio" package [41].

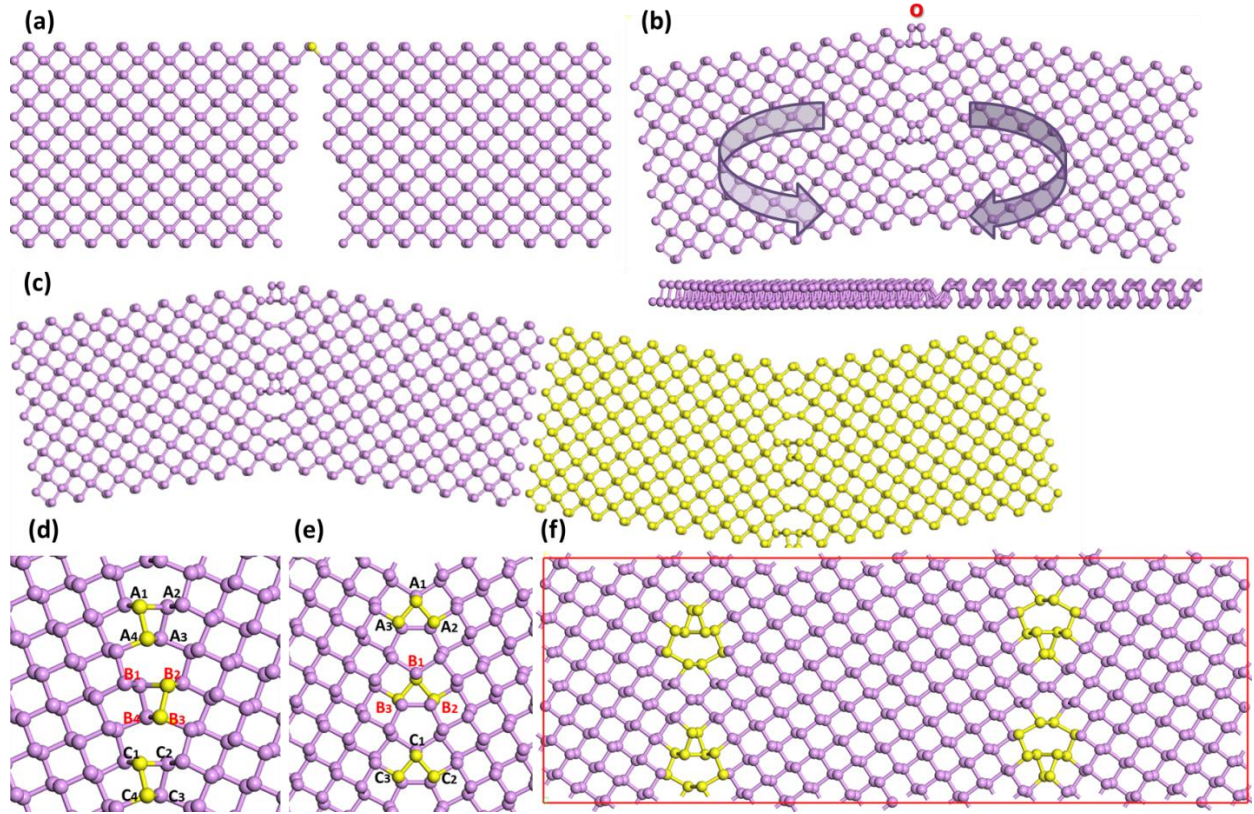

**Figure 1:** Construction of a phosphorene sample with vertically oriented grain boundaries: (a) Deletion of the selected atoms along the specific atomic rows. (b) Rotation of the two regions around the origin (O-atom) in the clock-wise and anti-clock-wise directions, respectively, and subsequent welding (top and side view). (c) Implementation of periodic boundary conditions in the direction perpendicular to the grain boundary. (d) Geometry of (4|8) defect pairs: the highlighted atoms ($A_1$, $A_4$, $B_2$, $B_3$, $C_1$, and $C_4$) are located in the upper plane of phosphorene. (e) Geometry of (5|7) defect pairs: the highlighted atoms ($A_1$, $A_2$, $A_3$, $C_1$, $C_2$, and $C_3$) are located in the upper plane of phosphorene, while ($B_1$, $B_2$, and $B_3$) in the lower plane. (f) Sample with two grain boundaries:

**each grain boundary contains two (4|8) defect pairs separated by two hexagons. The (4|8) defect pairs (edge dislocations) are oriented in opposite directions in these two grain boundaries. The red line outlines the computational box.**

We constructed a set of phosphorene samples with different tilt angles. The tilt angle (or misalignment angle) is defined as $\alpha = \alpha_L + \alpha_R$, where $\alpha_L$ and $\alpha_R$ are the angles between the crystallographic axis of the left and right grain domains of phosphorene. The tilt angle was varied by changing the number of hexagons separating the adjacent (4|8) or (5|7) defect pairs. A sample with the maximal tilt angle can be constructed by placing defect pairs as closely as possible ("head-to-tail") without the intervening hexagons. With introduction of hexagons between the successive defect pairs, the tilt angle is reduced. In our simulations, the number of hexagons introduced along the grain boundary between the adjacent defect pairs was varied from zero to five. The defect pairs (edge dislocations) were evenly distributed along the grain boundaries. The length of the grain boundaries was varied in the range from L≈30 Å to L≈80 Å, while the distance between the two grain boundaries is in the range from d≈30 Å to d≈60 Å. We constructed the samples containing only two defect pairs within the computational box (with exception for the sample with the highest tilt angle, where six defect pairs were hosted). For the samples with the low tilt angles, an increase in the length of the computational box along the grain boundary was compensated by the corresponding decrease in the distance between the grain boundaries. The total number of atoms used in our DFTB simulations was varied between ∼300 to ∼550 atoms.

Periodic boundary conditions were applied in all three directions. To avoid the self-interaction due to artificial periodicity along the Z direction, a vacuum slab with the width of *w*=30 Å was added in this direction. Initially, the geometry of the constructed samples was optimized, where both the atomic positions and the computational box shape were adjusted by minimizing the total energy with conjugate gradient method. Then we applied a uniform uniaxial tensile strain quasi-statically at zero temperature in the direction perpendicular to the grain boundaries. The tensile strain was increased gradually by a small step of $\delta\varepsilon$=0.01 until a failure strain was reached. At each step, we minimized the total energy. The self-consistent charge calculations were carried out at each step of the energy minimization. The k-point set for the Brillouin-zone integration was chosen by using the Monkhorst-Pack method [42]. The Monkhorst-Pack grid [42] with an 8x2x2 sampling set was adapted for Brillouin-zone integration. Following the previous DFTB studies for phosphorene nanoribbons and nanotubes [11,12,37,38], the s- and p-orbitals were specified for every P-atom. The Slater-Koster files [34] for phosphorus atoms were taken from 'MATSCI' set [43].

We also estimated the nominal stress, as explained in [9], and calculated and visualized the atomic strain along the direction of applied uniaxial strain using OVITO [44] software. In our failure analysis, we set the critical (maximal) bond length as $l_{max}$=2.95 Å since at this distance the attraction between a pair of phosphorous (P) atoms is insignificant. Thus a P-P bond can be strained up to ∼20% of its equilibrium value [45]. In our DFTB simulations, the equilibrium value $l_{eq}$=2.45 Å was obtained by optimizing the geometry of bulk phosphorene.

# 3. Results and Discussion

## 3.1 Geometry optimization

As the first step, we optimized the geometry of the constructed phosphorene samples with embedded grain boundaries by minimizing the total energy. Two samples are shown in Figure 2: a sample oriented along the AC direction with the grain boundaries composed of closely packed (4|8) rings (see Figure 2(a)) and a sample oriented along the ZZ direction with the grain boundaries composed of closely packed (5|7) rings (see Figure 2(b)). The defect rings are highlighted in Figure 2(a, b).

The most notable feature of these free-standing phosphorene samples with optimized geometry is their planar structure. It is well-known that introduction of grain boundaries in free-standing graphene leads to out-or-plane warping near the grain boundaries [14]. These out-of-plane corrugations act as an efficient mechanism for relieving the in-plane strain at grain boundaries in graphene. They effectively 'screen' the in-plane elastic fields produced by edge dislocations [14,23,46]. In some cases, with specific arrangements of the pentagons and heptagons at grain boundaries, graphene becomes inflected (with inflection angle up to ~72°), which significantly reduces the mechanical strength of graphene [47]. Similar inflection was also found in hexagonal boron-nitride monolayers with specific grain boundaries [17]. In contrast, phosphorene with grain boundaries sustains its original planar shape, without wrinkles, inflections and out-of-plane buckling due to its puckered structure. Its lattice structure, defined by tetrahedral geometry of sp$^3$ hybridized P-P bonds, is very flexible. We examined the bond angles in 4-, 5-, 7- and 8-member rings of defect pairs constituting grain boundaries. It was found that the values of these bond angles (82°-117°) are very close to the original ones of pristine phosphorene (92°–108°).

In addition, we calculated the energy per unit length of grain boundary $E_{gb}$ according to:

$$E_{gb} = \frac{E_{tot} - NE_b}{2L}$$

where $E_{tot}$ is the total energy of a sample containing two grain boundaries, $N$ is the total number of atoms, $2L$ is the total length of the two grain boundaries, and $E_b$ is the energy per atom ($E_b = 27.21 eV$) in pristine (defect-free) phosphorene. The energy per unit length as a function of tilt angle is shown in Figure 2(c) for the phosphorene samples containing grain boundaries composed of (4|8) defect pairs (red circles) and (5|7) defect pairs (blue squares). The GB energies for the two types of grain boundaries are quite similar. Nonetheless, the grain boundaries formed by (5|7) defects have slightly lower energy, and therefore they are slightly more stable. Similar to graphene [14] and hexagonal boron-nitride [17], grain boundary energy in phosphorene increases with an increase in the tilt angle (while the tilt angle is less than 60° due to six-fold symmetry of hexagonal lattice).

The range of the grain boundary energy per unit length in phosphorene is lower than that of graphene ($max\ E_{gb} \sim 0.8\ eV/Å$) [14], boron-nitride ($max\ E_{gb} \sim 1\ eV/Å$) [17] and MoS$_2$ ($max\ E_{gb} \sim 0.7\ eV/Å$) [40]. The smaller GB energies of phosphorene are due to its flexible lattice. In contrast to sp$^2$ hybridized C-C bonds of graphene, sp$^3$ hybridized P-P bonds of phosphorene constitute an especially malleable lattice structure, accommodating more dislocations than other 2D materials, particularly along the AC direction [30]. The low grain boundary energy also signifies that these GBs can be formed relatively more easily than those of graphene and boron-nitride.

We also calculated Young's modulus of phosphorene with embedded grain boundaries by subjecting it to small (less than 5%) tensile and compressive uniaxial strains. In Figure 2(d), we plot the obtained Young's modulus for grain boundaries composed of (4|8) defects (red circles) and (5|7) defects (blue squares) as a function of tilt angle. It can be seen that the Young's modulus increases only marginally with the tilt angle for both types of grain boundaries.

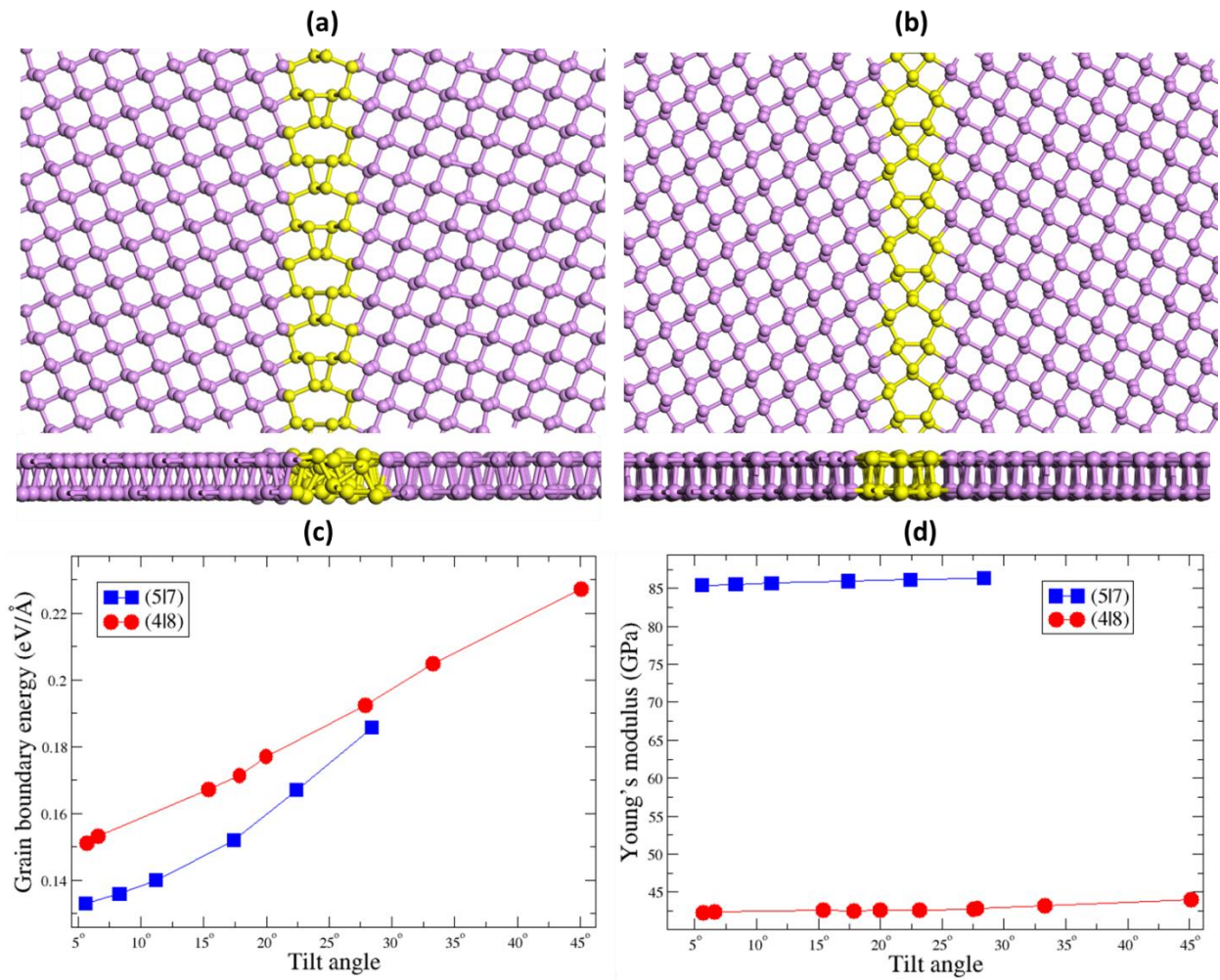

Figure 2: (a) Grain boundary formed by an array of the closely spaced (4|8) defect pairs (highlighted by yellow) between two AC-oriented phosphorene grains (top and side views). (b) Grain boundary formed by an array of the narrowly spaced (5|7) defects (highlighted by yellow) between two ZZ-oriented phosphorene grains (top and side views). (c) Energy of grain boundary (per unit length) formed by (4|8) defect pairs (red circles) and (5|7) defect pairs (blue squares) as a function of tilt angle. (d) Young's modulus as a function of tilt angle for phosphorene samples containing grain boundaries composed of (4|8) dislocations (red circles) and (5|7) dislocations (blue squares).

We also examined the energy distribution around the constructed grain boundaries, using DFTB+ capacity to calculate energy per atom. In Figure 3(a-c), a color map is used to indicate the energy per atom in the AC oriented phosphorene sample containing a grain boundary formed by (4|8) dislocations. Three samples with a low $\alpha=6.6°$ (a), an intermediate $\alpha=20.1°$ (b) and the highest $\alpha=45.0°$ (c) tilt angles are shown. The largest variation of the energy is observed at the grain boundary, while away from it, the energy rapidly converges to the average bulk value. The abrupt convergence of the energy with the distance from the grain boundary matches the well-known exponential decrease in strain energy near GBs [14].

The atoms constituting (4|8) defect rings have the highest energy, especially the pairs shared by the octagon (marked by red) and the adjacent hexagons (see Figure 3(a, b)). Later on, we will show that these pairs (and the most stretched P-P bonds connecting them) play a crucial role in tensile failure of phosphorene with GBs. The significant energy deviation from the average bulk value is typical only for the atoms of (4|8) defects. The energy of the hexagon atoms located in between the consecutive (4|8) defects at the grain boundary is noticeably lower and closer to the average bulk value.

Similarly, in Figure 3(d-f), we show the energy distribution in the ZZ oriented sample containing a grain boundary formed by (5|7) defects. Three samples with a low $\alpha=5.6°$ (d), an intermediate $\alpha=11.2°$ (e), and the highest $\alpha=28.4°$ (f) tilt angles are presented. Similar to the previous case, the energies of the atoms in (5|7) defects are the highest ones. The energies of the hexagon atoms surrounding the (5|7) defects are only to some extent larger than those in the bulk. The highest energy is attained by the pairs of heptagon atoms (red atoms in Figure 3(d-f)) of the (5|7) defects sharing the bonds with the nearest pentagon atoms. As will be discussed later, these atoms and their connecting bonds play a critical in tensile failure of the samples.

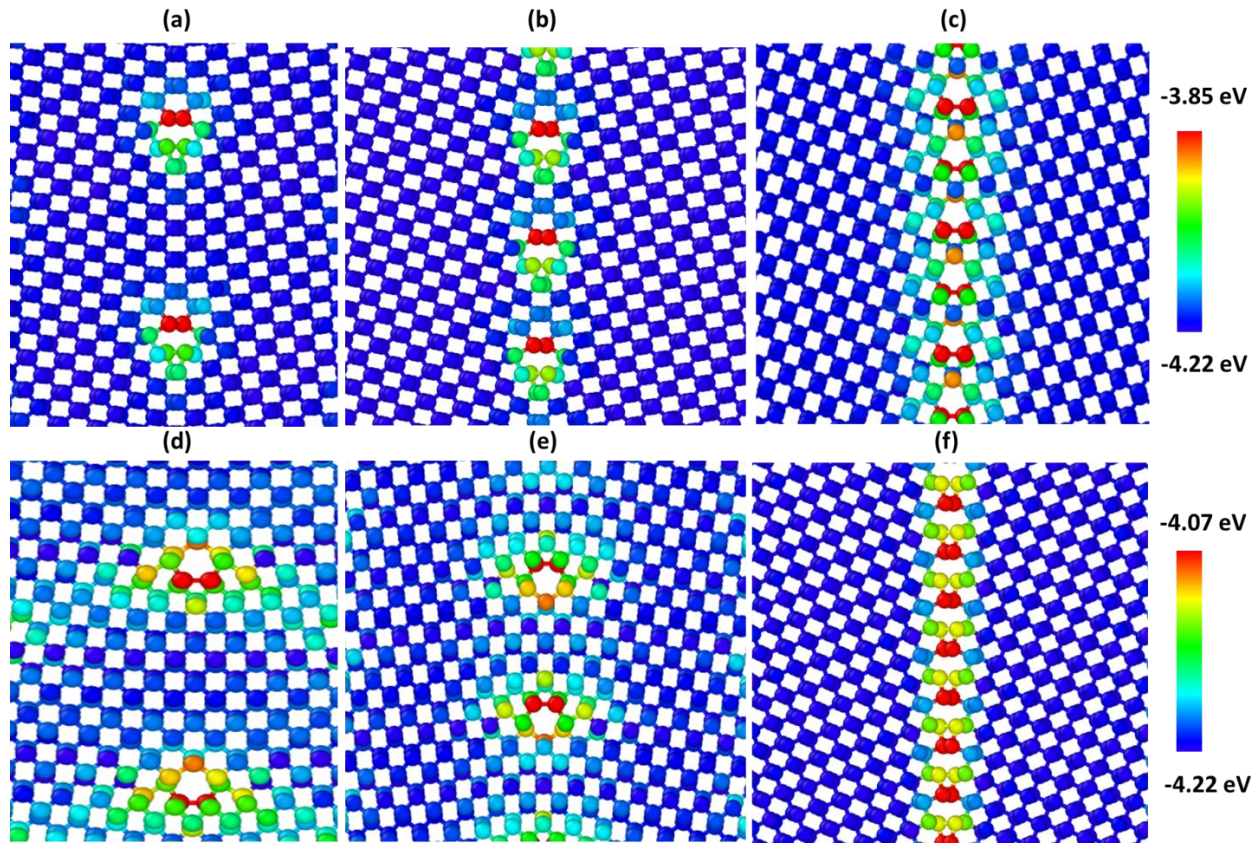

Figure 3: (a-c) Energy distribution (energy per atom) in the AC oriented phosphorene with the grain boundary formed by (4|8) dislocations. The samples with the low α=6.6° (a), intermediate α=20.1° (b) and high α=45.0° (c) tilt angles. (d-f) Energy distribution (energy per atom) in the ZZ oriented phosphorene with the grain boundary formed by (5|7) dislocations. The samples with the low α=5.6° (d), intermediate α=11.2° (f) and high α=28.4° (e) tilt angles. The highest and the lowest values of the energy per atom are indicated at the color bars.

### 3.2 Tensile deformation: energy

To begin with, we apply a uniaxial tensile strain by quasi-statically stretching phosphorene samples in the direction perpendicular to the grain boundary. In Figure 4, we plot the energy (per atom) in the phosphorene samples containing the grain boundary formed by (4|8) defects at the different values of tensile strain. The displayed snapshots (see Figure 4(a-c)) are for the samples with a low, an intermediate and the highest tilt angles, respectively. The side color bars indicate the lowest and highest values of energy per atom. It is seen that the energy per atom increases with the tensile strain and, in particular, the energy of the grain boundary atoms. As can be seen in Figure 4, the increase in the atom energy is different for the samples with the low and high tilt angles. Clearly, an increase in the number of hexagons separating the (4|8) defects leads to significantly different responses to the applied uniaxial tensile strain. The energy distributions for samples with the low and the intermediate tilt angles are shown in Figure 4(a-f). As discussed above, the energy distribution is non-uniform: the energy of the (4|8) ring atoms at the grain boundary is markedly above the average, with the highest energy attained by the atoms sharing the bonds between the octagons and the adjacent hexagons. When the applied tensile strain increases, the energy of these atoms rises progressively faster than the energy of the remaining ones. Once a failure strain is reached, the bonds between these atoms rupture (see Figure 4(d, f)).

We found that the phosphorene sample with the highest tilt angle GB, composed of the closely ("head-to-tail") packed (4|8) defects behaves differently under the applied tensile strain (see Figure 4(g-i)). The sample can be significantly stretched along the AC direction ($\varepsilon \approx 0.41$), which is considerably larger than any sample with low-to-intermediate value of tilt angle ($\varepsilon \approx 0.2$-$0.3$), but noticeably smaller than pristine phosphorene ($\varepsilon \approx 0.5$). As can be seen in Figure 4 (g-i), the atoms of the (4|8) defects have the highest energy, which increases with applied tensile strain. However, the difference in the energy between the GB and bulk atoms is rather smaller for the highest tilt angle sample in comparison with the lower tilt angle samples.

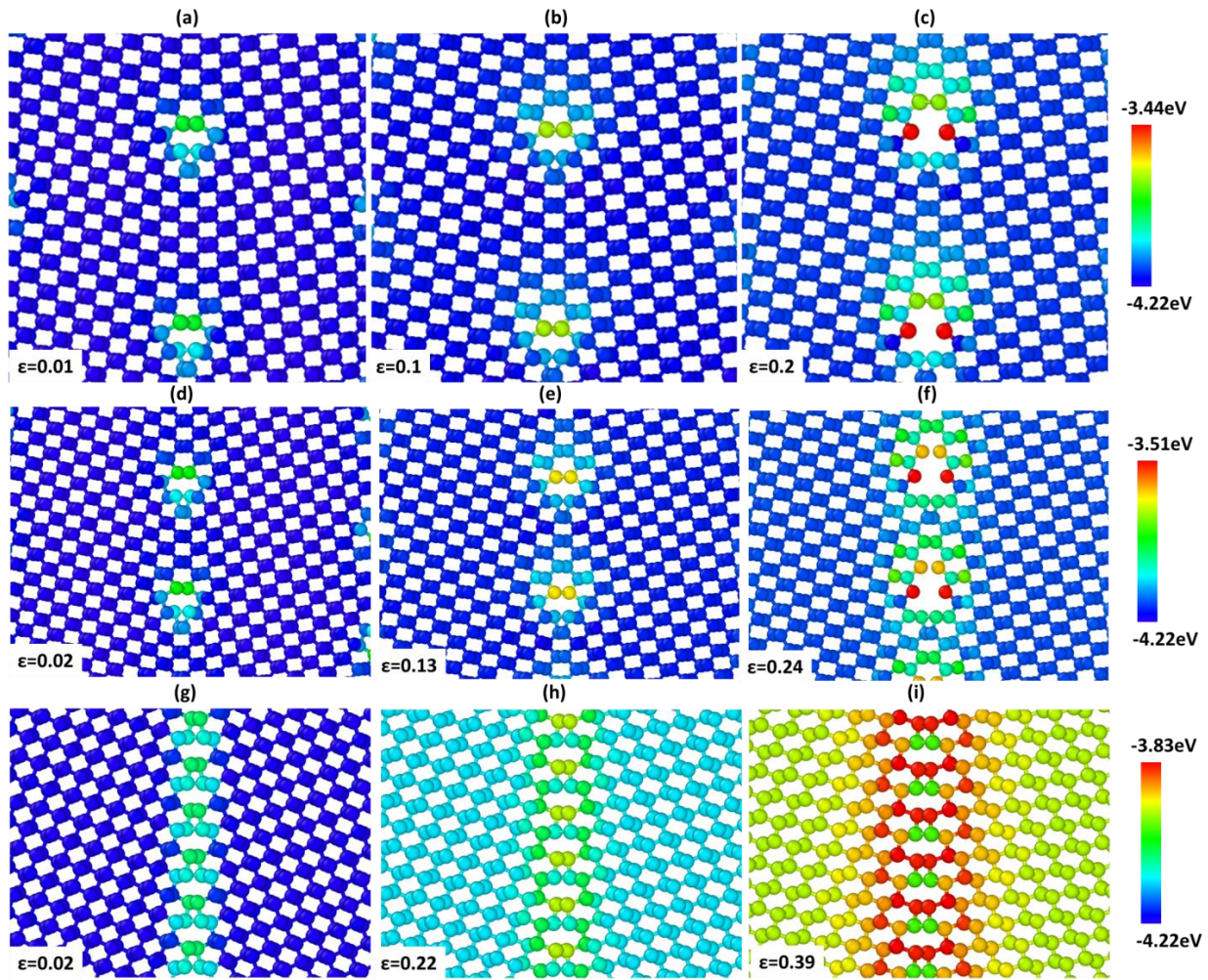

**Figure 4: Distribution of energy per atom in the AC oriented phosphorene samples with the grain boundaries formed by (4|8) defect pairs under applied uniaxial tensile strain. Snapshots (a-c): the sample with the low tilt angle α=6.6° at ε = 0.01 (a), ε = 0.1 (b) and ε = 0.2 (c). Snapshots (d-f): the sample with the intermediate tilt angle α=20.0° at ε = 0.02 (d), ε = 0.13 (e) and ε = 0.24 (f). Snapshots (g-i): the sample with the highest tilt angle**

In Figure 5, we illustrate the effect of applied tensile strain on the energy distribution for the phosphorene sample elongated along the ZZ direction. The phosphorene samples with a low (α=8.3º, see Figure 5 (a-c)) and high (α=28.4º, see Figure 5 (d-f)) tilt angles are shown. Like the previous case, the energy per atom rises with the applied tensile strain, and the largest increase occurs for the (5|7) ring atoms at the grain boundary. As can be seen in Figure 5, the heptagon atoms of the (5|7) defects connected to the adjacent hexagons alongside the grain boundary gain the highest energy. The bonds connecting these atoms to their nearest in-plane neighbors are the most strained bonds in the phosphorene sample at any given strain. Tensile failure begins when these most stretched bonds start to break (see Figure 5). The phosphorene sample with the highest tilt angle (α=28.4º) can be stretched by ε≈0.18 along the ZZ direction, which is larger than that with the low tilt angle (ε≈0.14), but is notably smaller than pristine phosphorene (ε=0.25).

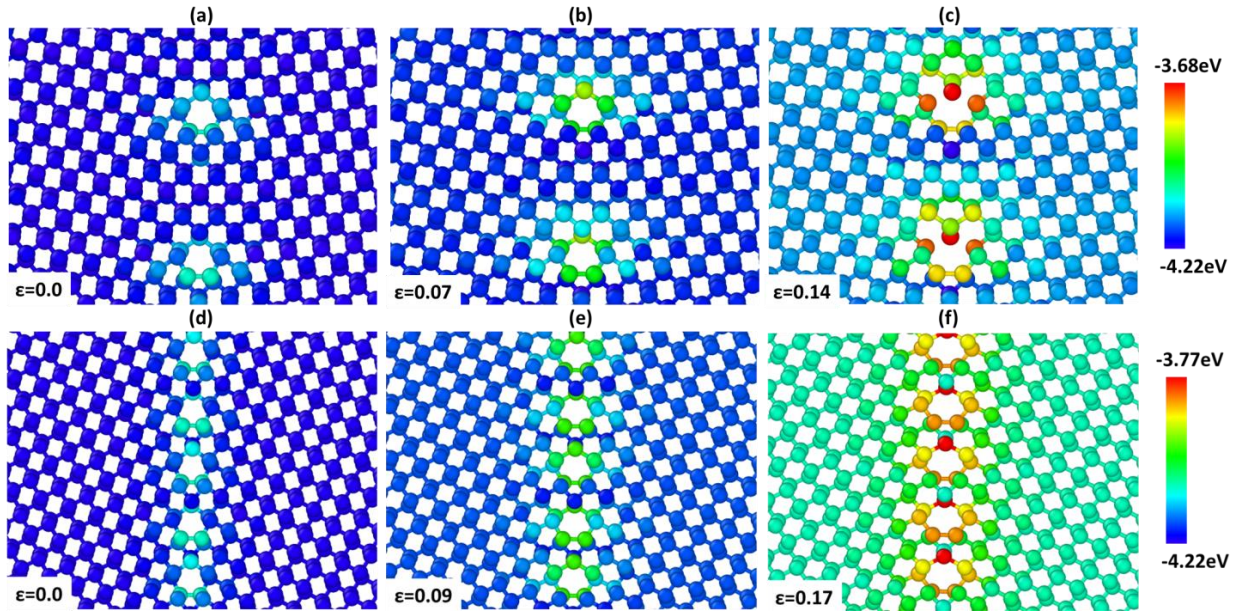

Figure 5: Distribution of energy per atom in the ZZ oriented phosphorene samples with the grain boundaries formed by (5|7) dislocations under applied uniaxial tensile strain. Snapshots (a-c): the sample with the low tilt angle α=11.2º at ε = 0.0 (a), ε = 0.07(b) and ε = 0.14 (c). Snapshots (d-f): he sample with the highest tilt angle α=28.4º at ε = 0.0 (d), ε = 0.09 (e) and ε = 0.17 (f). The highest and the lowest values of the energy per atom are indicated at the color bars.

### 3.3 Tensile deformation: atomic strain

Next, we examined the atomic strain distribution along the direction of applied uniaxial tensile strain. Figure 6 plots the atomic strain distribution for a phosphorene sample containing the GB formed by (4|8) edge dislocations. In snapshots (a-c) of Figure 6, we map the atomic strain distribution of the sample with a low tilt angle α=6.6º at different values of applied strain. Snapshots (d-f) show for the sample with an intermediate tilt angle α=20.0º, while the snapshots (g-i) illustrate strain distribution of the sample with the highest α=45.0º tilt angle. The maximal value of atomic strain is indicated at the color bars. It can be seen that at the initial stage of the deformation, when the applied uniaxial strain is relatively small, the atomic strain is evenly distributed throughout the samples: in Figure 6 (a, d, and g), all the atoms have the same color. When the applied strain rises, the initial uniformity of the atomic strain is disrupted once the highest magnitudes of the atomic strain are attained by the some grain boundary atoms (see Figure 6). More specifically, the maximal atomic strain is reached by the (4|8) defect atoms sharing the bonds between the octagons and the adjacent hexagons alongside the grain boundary.

The variation in the atomic strain distribution is particularly different for the phosphorene sample with the highest tilt angle (see Figure 6 (g-i)). In this sample, the atomic strain of the (4|8) defect atoms is smaller than that of the remaining atoms. The difference becomes more and more apparent with an increase in applied tensile strain. We found that this is due to the mutual cancellation of the tensile and compressive stress fields appearing around the (4|8) defects since the 4-membered rings create compressive stress fields, while the 8-membered rings produce tensile stress fields. Because these defects are closely spaced in the 'head-to-tail' fashion, their stress fields partially cancel each other, causing the reduction in the atomic strain.

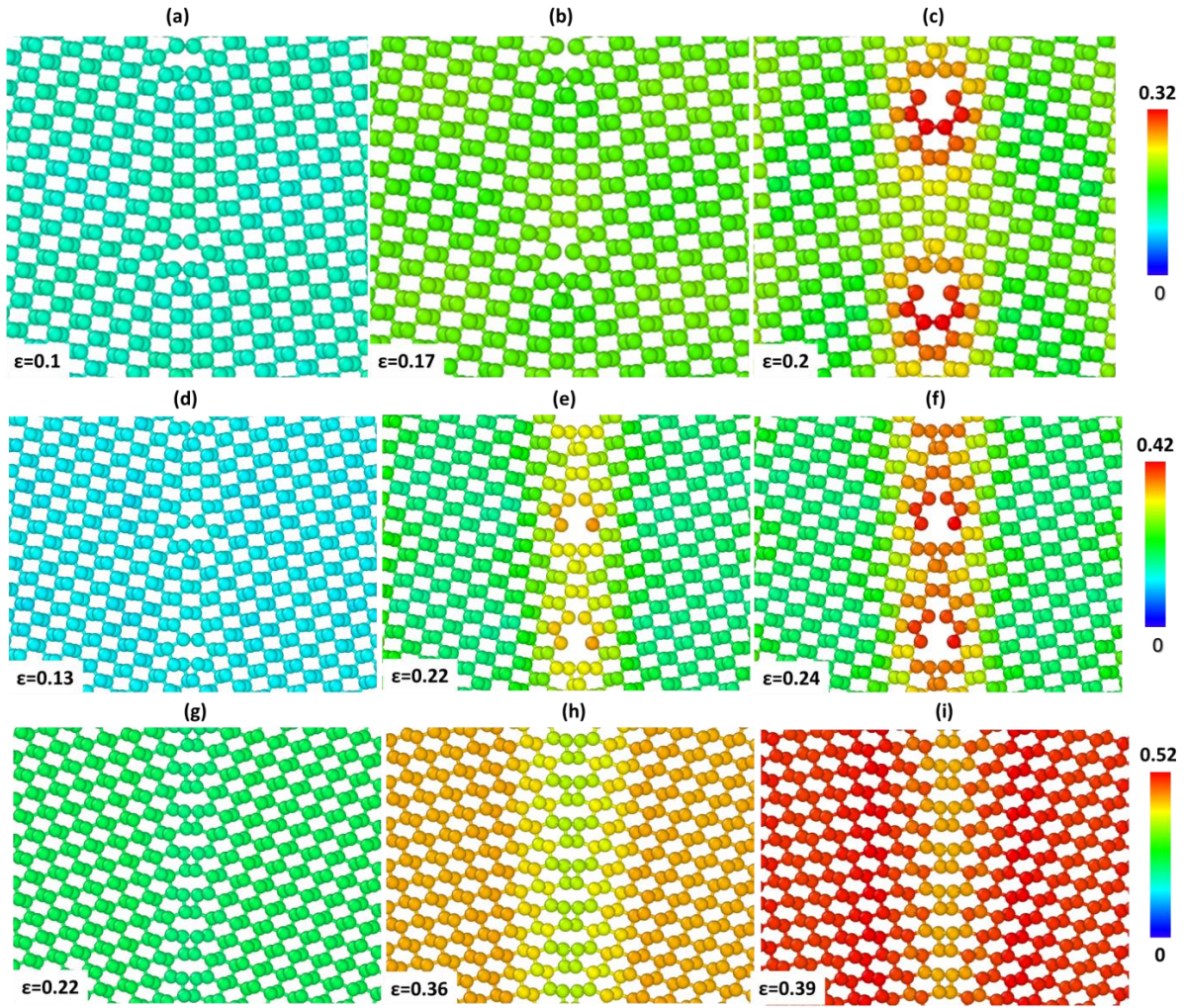

Figure 6: Atomic strain distribution in the phosphorene samples with the grain boundaries formed by (4|8) defect pairs under tensile strain. Snapshots (a-c): the sample with the low tilt angle α=6.6° at ε = 0.1 (a), ε = 0.17(b) and ε = 0.2 (c). Snapshots (d-f): the sample with the intermediate tilt angle α=20.0° at ε = 0.13 (d), ε = 0.22 (e) and ε = 0.24 (f). Snapshots (g-i): the sample with the highest tilt angle α=45.0° subjected to tensile strain at ε = 0.22 (g), ε = 0.36 (h) and ε = 0.39 (i). The color bars with the indicated maximal value of atomic strain are shown.

Variation of the atomic strain in phosphorene oriented along the ZZ direction subjected to uniaxial tensile strain is shown in Figure 7. We illustrate the atomic strain distribution in the samples with the low (α=11.2°, see Figure 7 (a-c)) and high (α=28.4°, see Figure 7 (d-f)) tilt angles at three different values of applied strain. The atomic strain in the samples with the grain boundaries composed by (5|7) defects varies in the following way: Initially, when the applied uniaxial strain is rather small, the atomic strain is uniform, as indicated by the same mapping color in Figure 7 (a, d). When the applied tensile strain increases further, approaching a critical value of failure strain, the non-uniformity becomes visible in the atomic strain distribution, especially at the grain boundary. Near the critical strain, the highest values of the atomic strain are attained by the atoms of the (5|7) defects (see Figure 7). The maximal value of the atomic strain is acquired by the heptagon atoms of the (5|7) defect rings sharing the bonds with the adjacent hexagon atoms.

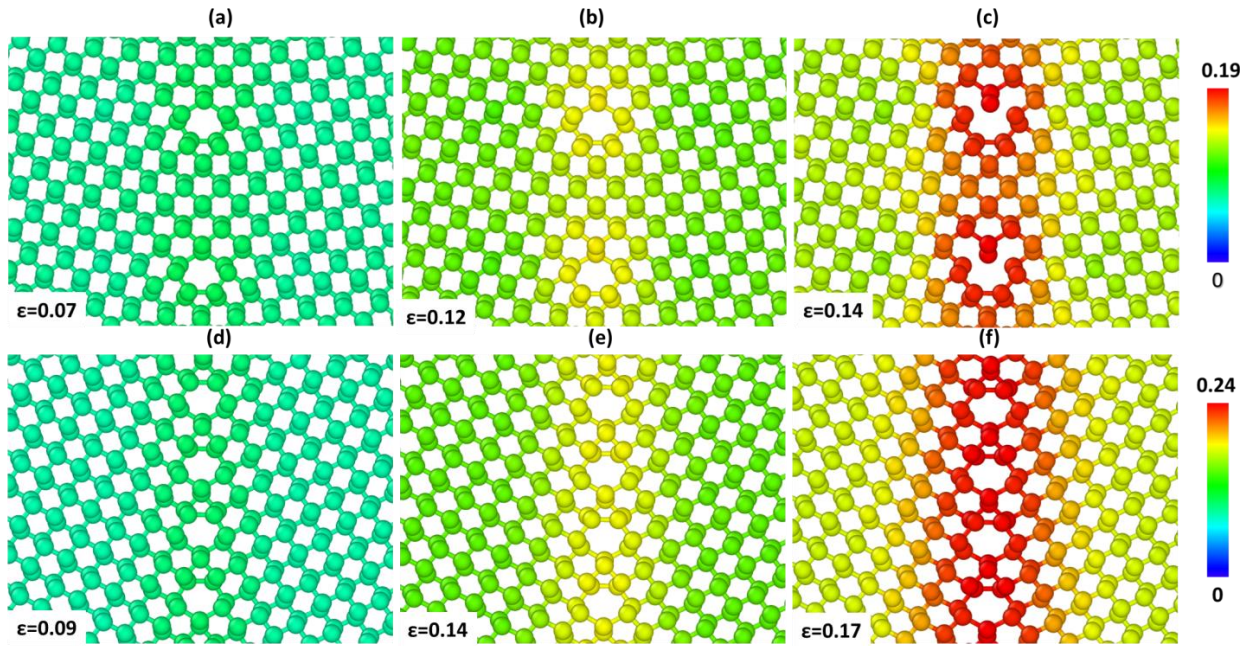

Figure 7: Atomic strain distributions in the phosphorene samples with the grain boundaries formed by (5|7) defect pairs under applied uniaxial tensile strain. Snapshots (a-c): the sample with the low tilt angle α=11.2° at ε = 0.07 (a), ε = 0.12 (b) and ε = 0.14 (c). Snapshots (d-f): the sample with the high tilt angle α=28.4° at ε = 0.09 (d), ε = 0.14 (e) and ε = 0.17 (f). The color bars with the indicated maximal values of atomic strain are shown.

## 3.4 Failure mechanism

Next, we examined the failure mechanism: a series of atomic-scale events that lead to ultimate failure. Figure 8 shows the initial stage of tensile failure of the AC oriented (see Figure 8(a-c)) and ZZ oriented (see Figure 8(d-e)) phosphorene samples with GBs. The first critical bonds to rupture (see highlighted bonds in Figure 8) are the most stretched bonds in the octagons of (4|8) dislocations and heptagons of (5|7) dislocations. The critical bonds are oriented along the direction of applied tensile strain. These are the bonds separating the octagons and the adjacent hexagons in (4|8) defects, and the pentagon and the adjacent hexagons in (5|7) defects (similar to graphene [22,24,48] and h-BN [17], where fracture begins with a shared heptagon-hexagon B–N bond). As mentioned above, a single (4|8) defect pair is an edge dislocation with the compression stress on the 4-side and tension stress on the 8-side of the defect rings. In the adjacent 4-6 and 8–6 pairs, the shared P-P bonds also experience compression on the 4-side and tension on the 8-side. Likewise, a single (5|7) defect is an edge dislocation with the compression stress on the 5-side and tension stress on the 7-side of the defect rings. Thus in the neighboring 5–6 and 7-6 pairs, the shared P-P bonds undergo compression on the 5-side and tension on the 8-side. This initial tensile pre-strain of the shared 8-6 and 7-6 P-P bonds of the octagons and heptagons can be further increased by applied tensile strain. Therefore, the first critical bonds which break under additional tension must be these most intrinsically stretched ones separating the octagons (or heptagons) and the adjacent hexagons (see Figure 8(a-b, d-f)). After the rupture of the first critical bond, a next neighboring bond in the same hexagonal ring is stretched and broken, and then the complete tensile failure promptly follows along the grain boundary. This observation is universal for all the samples that we considered here; except for the samples with the highest tilt angles, where the defect density is the highest. In the two samples with the highest tilt angles (one with GB formed by (4|8) dislocations and the other by (5|7) dislocations), tensile failure occurs in a different way as evident in Figure 8(c). It can be seen that the closely connected (4|8) edge dislocations of the ZZ oriented sample remain intact, while the side bonds linking them to the neighboring hexagons rupture. The similar observation is valid for the closely 'head-to-tail' packed (5|7) edge dislocations forming in the AC-oriented sample (see Figure 8(f)).

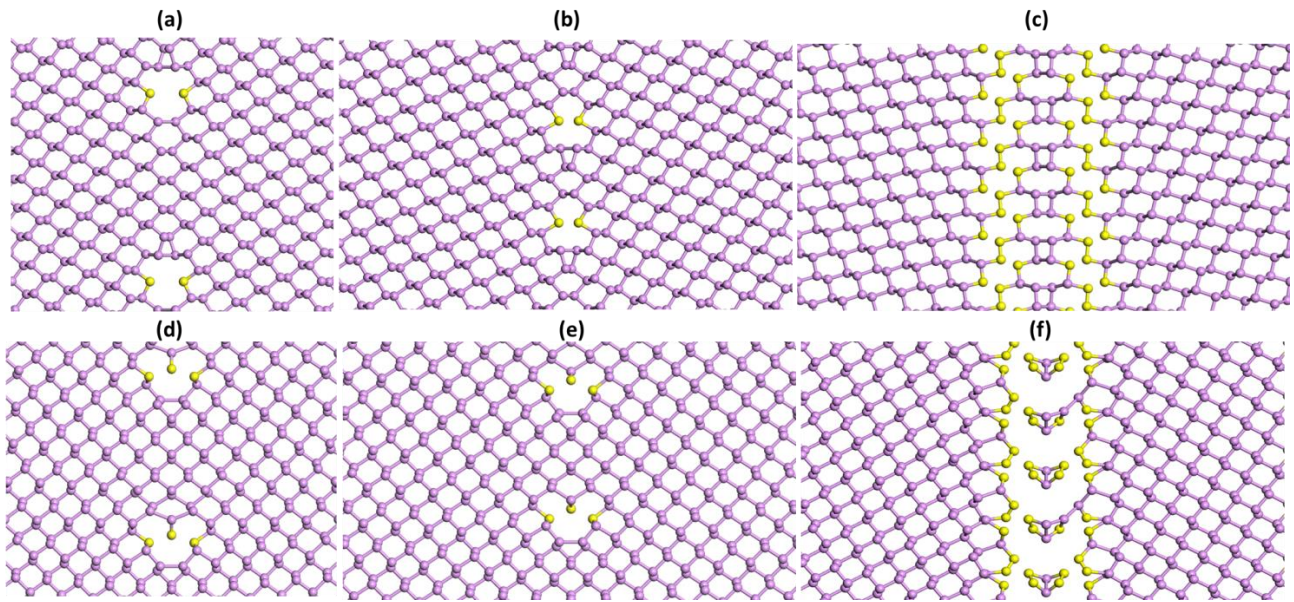

Figure 8: Failure initiation in phosphorene with linear grain boundaries: (a-c) Failure of the AC oriented phosphorene with the grain boundary formed by (4|8) dislocations. The samples with the low α=6.6° (a), intermediate α=20.0° (b) and highest α=45.0° (c) tilt angles are shown. (d-f) Failure of the ZZ oriented phosphorene with the grain boundary formed by (5|7) dislocations. The samples with the low α=8.3° (d),

**intermediate α=11.2° (e) and highest α=28.4°(f) tilt angles are shown. Failure is initiated as the most stretched octagon-hexagon bonds of the (4|8) dislocations and heptagon-hexagon bonds of the (5|7) dislocations rupture. The P-atoms with broken bonds are highlighted by yellow.**

Figure 9 shows the critical bonds ("weakest links"), that is, heptagon-hexagon bonds of (5|7) defect pair (see highlighted $A_1A_2$ bonds in Figure 9 (a)) and octagon-hexagon bond of (4|8) defect pair (see highlighted $A_1A_1$ bond in Figure 9 (b)). As discussed above, these "weakest links" rupture first at the outset of failure (see Figure 8). In order to understand why these bonds break first at lower tensile strain in the samples with low tilt angles, we measured the initial length of these bonds in the samples with the different tilt angles. Figure 9(c) plots the initial bond length of the octagon-hexagon bonds of (4|8) defect pairs (red circles) and the heptagon-hexagon bonds of (5|7) defect pairs (blue squares) as a function of tilt angle at zero strain. As can be seen in Figure 9 (c), the smaller the tilt angle, the larger the initial length (and the initial pre-strain) of the critical bonds. Conversely, the larger the tilt angle, the closer the (4|8) or (5|7) edge dislocations to each other, and the more substantial the mutual cancellation of their overlapping compressive and tensile stress fields. Hence, the large tilt-angle GBs are better off in accommodating edge dislocations by reducing the pre-strain of the critical bonds. The reduced pre-strain leads to the increase in the failure strain and the failure stress (intrinsic strength) of phosphorene with grain boundaries.

We also studied how the length of the selected bonds changes under applied tensile strain (see Figure 9 (d, e)). The length of the critical octagon-hexagon bonds of the (4|8) defects as a function of applied tensile strain is plotted in Figure 9(d) for the samples with different tilt angles: α=5.7° (magenta circles), α=6.6° (yellow circles), α=15.4° (blue circles), α=20.0° (green circles), α=27.9° (red circles) and α=45.0° (black circles). There is an evident difference between the sample with the highest tilt angle and the remaining ones. Introduction of a single hexagon between two consecutive (4|8) rings leads to a drastic change in the bond length. The length of the critical bonds in the sample with the highest tilt angle increases monotonically with applied strain. In all other samples with lower tilt angles, the bond length increases non-monotonically up to the failure strain.

Figure 9(e) plots the bond length of the critical heptagon-hexagon bonds of the (5|7) defects as a function of applied tensile strain. The bond length is calculated for the samples with various tilt angles: α=5.3° (yellow circles), α=8.3° (blue circles), α=11.2° (green circles), α=17.4° (red circles) and α=28.4° (black circles). In contrast to the critical bonds of the (4|8) defects, the length of the critical bonds of the (5|7) defects increases almost linearly with strain until the failure point is reached.

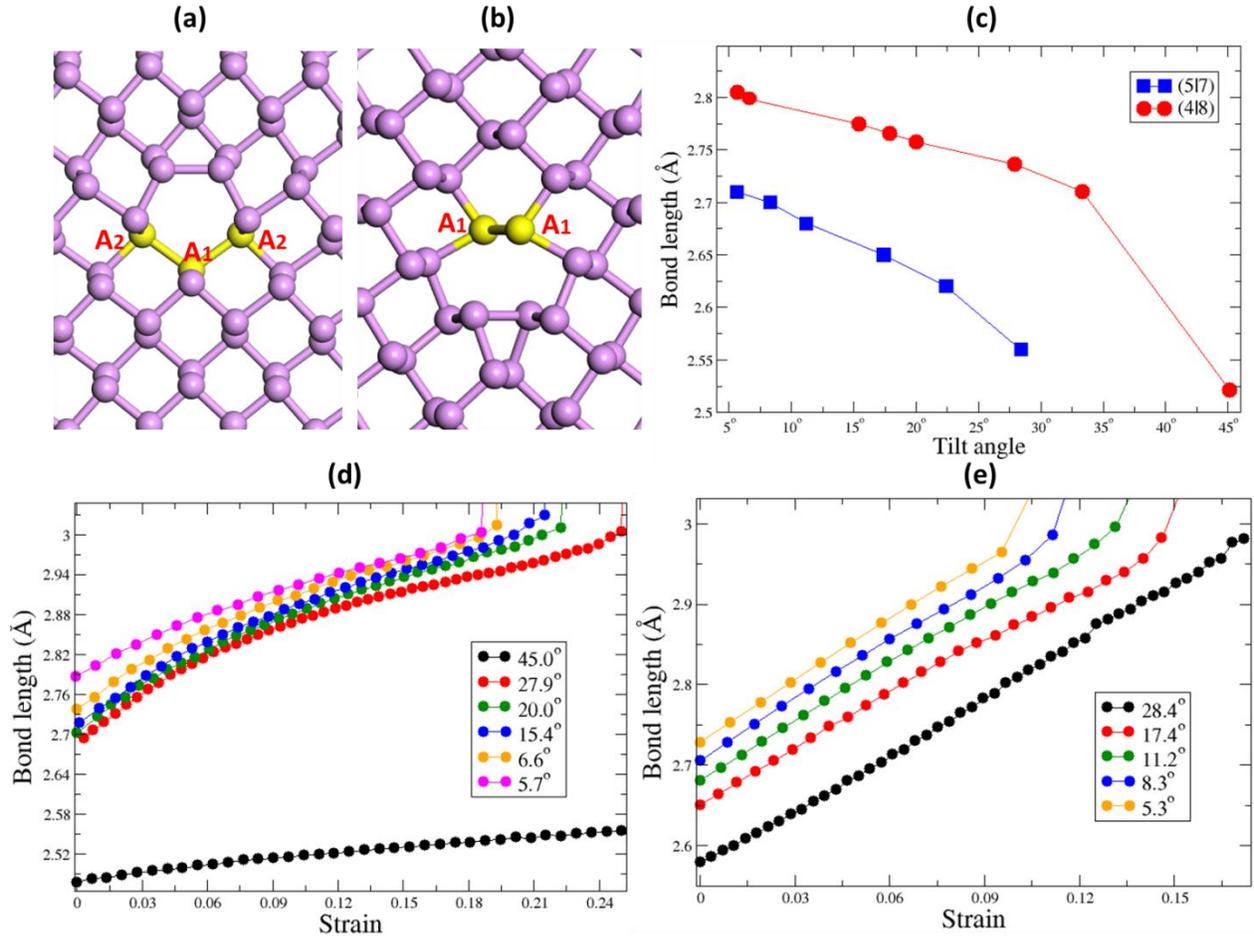

**Figure 9:** The "weakest links" (a-b): (a) the highlighted critical bonds $|A_1A_2|$ in heptagon of a (5|7) defect pair and (b) highlighted critical bonds $|A_1A_1|$ in octagon of a (4|8) defect pair rupture first at the onset of tensile failure. (c) Initial bond length of the critical $|A_1A_1|$ bonds in (4|8) defect pairs (red circles) and of the $|A_1A_2|$ bonds in (5|7) defect pairs (blue squares) as a function of tilt angle at zero strain. The smaller is the tilt angle, the larger is the initial pre-strain of the critical bonds. (d) Bond length of the critical $|A_1A_1|$ bonds in (4|8) dislocations as a function of applied uniaxial tensile strain. The bond length is calculated for the samples with different tilt angles: α=5.7° (magenta circles), α=6.6° (yellow circles), α=15.4° (blue circles), α=20.0° (green circles), α=27.9° (red circles) and α=45.0° (black circles). (f) Bond length of the critical $|A_1A_2|$ bonds in (5|7) dislocations as a function of applied uniaxial tensile strain. The bond length is calculated for the samples with various tilt angles: α=5.3° (yellow circles), α=8.3° (blue circles), α=11.2° (green circles), α=17.4° (red circles) and α=28.4° (black circles).

It is interesting to compare initial pre-strain of the critical bonds before the application of external strain in phosphorene and graphene. The pre-strain of the critical bonds in ZZ oriented phosphorene and graphene with different tilt angles are given in Table 1. The similarity in the pre-strain values of critical bonds is evident.

**Table 1:** Pre-strain of critical bonds in ZZ oriented phosphorene and graphene[22] for different tilt angles.

| Tilt angle | 5.6° | 11.2° | 22.4° |
|---|---|---|---|
| Bond strain (phosphorene) | 9.1% | 7.8% | 5.5% |

| Tilt angle            | 5.5°  | 13.2° | 21.7° |
|-----------------------|-------|-------|-------|
| Bond strain (graphene)| 9.5%  | 8.7%  | 5.4%  |

The pre-strain of the critical bonds in AC oriented phosphorene and graphene with different tilt angles are given in Table 2. The noticeable difference in the pre-strain values of critical bonds is obvious: the bonds of phosphorene are more pre-strained than those of graphene.

Table 2: Pre-strain of critical bonds in AC oriented phosphorene and graphene[22] for different tilt angles.

| Tilt angle                | 15.4° | 20.01°| 27.8° |
|---------------------------|-------|-------|-------|
| Bond strain (phosphorene) | 17.2% | 16.1% | 15.4% |
| Tilt angle                | 15.8° | 21.4° | 28.7° |
| Bond strain (graphene)    | 23.4% | 9.3%  | 1.7%  |

## 3.5 Critical strain and stress

Figure 10 plots the failure strain (see Figure 10(a)) and the failure stress (see Figure 10(b)) of phosphorene with GBs formed by (4|8) dislocations (red circles) and (5|7) dislocations (blue squares) as a function of tilt angle. It is seen that both the failure strain and failure stress increase with tilt angle in both cases, indicating that the phosphorene samples with high-angle grain boundaries are more resilient than those with low-angle GBs. Since the rupture of these highly-strained critical bonds leads to tensile failure of phosphorene with GBs, it is the initial pre-strain of the critical bonds that governs the tensile failure and determines the failure strain and stress. As in the case of graphene [14,23,46,47,49], the obtained results cannot be explained by Griffith-type fracture mechanics criterion, because it does not take into account rupture of the pre-strained critical bonds, which is responsible for the tensile failure.

It must be noted that the failure strain for the phosphorene samples with grain boundaries composed of (4|8) defect pairs is higher than those of composed of (5|7) defects. It is known that phosphorene is mechanically anisotropic material [2], with the failure strain along the AC direction being much larger than that along the ZZ direction. This mechanical anisotropy is attributed to the puckered structure of phosphorene along the AC direction [2]. In contrast to the failure strain, the failure stress is considerably larger along the ZZ direction [2,50,51]. Consequently, this mechanical anisotropy in the failure strain and failure stress is preserved even with grain boundaries being embedded in phosphorene.

Our calculations show that the failure strain and failure stress along the AC direction for pristine phosphorene are $\varepsilon_{cr}$=0.49 and $\sigma_{cr}$=6.4 GPa, respectively. Clearly, the sample with the lowest tilt angle grain boundaries (constructed in our simulations) shows a maximal reduction in the failure strain by 61.2% and the failure stress by 46.9%. The sample with the highest tilt angle grain boundaries shows a minimal reduction of the failure strain by 16.3% and the failure stress by 15.1%. Obviously, the introduction of the grain boundaries considerably reduces the failure strain and failure stress of phosphorene compared to pristine (defect-free) one, and the failure strain and stress reduction depends on the GB tilt angle.

Our calculations show that the failure strain and the failure stress along the ZZ direction for perfect phosphorene are $\varepsilon_{cr}=0.25$ and $\sigma_{cr}=13.2$ GPa, respectively. Evidently, the sample containing the grain boundaries with the lowest tilt angle shows the maximal reduction in the failure strain by 52% and the failure stress by 20.9%. While the sample containing the highest GB tilt angle shows a minimal reduction in the failure strain by 28% and the failure stress by 12.9%. Apparently, the samples with large-angle grain boundaries are stronger than those with low-angle grain boundaries because the former is able to better accommodate the strained rings of the edge dislocations that the latter.

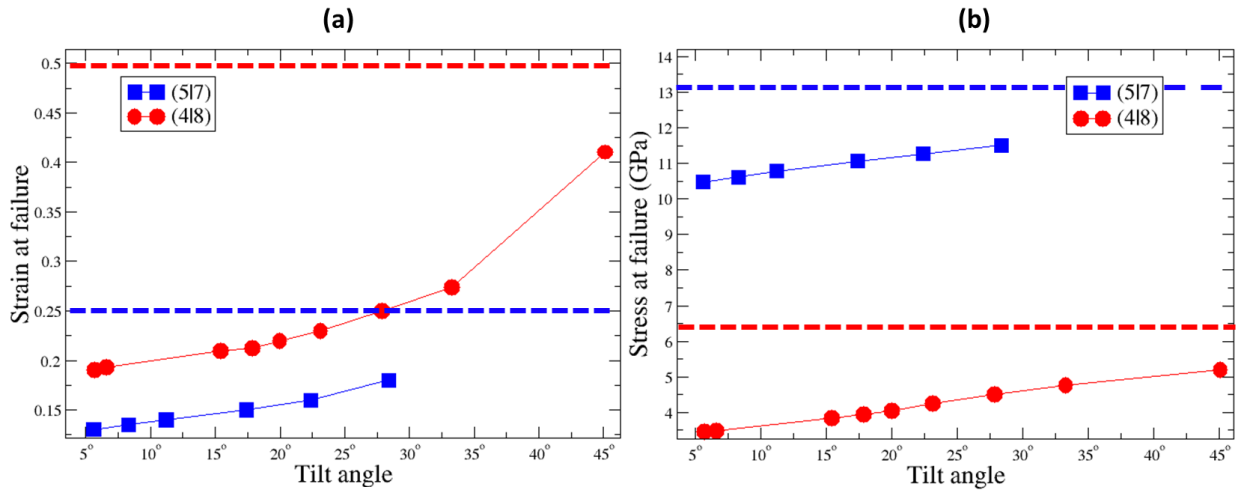

Figure 10: Failure strain (a) and failure stress (b) as a function of tilt angle for phosphorene with linear grain boundaries formed by evenly spaced (4|8, red circles) and (5|7, blue squares) edge dislocations. The dashed lines on the left panel (a) indicate the failure strain for pristine phosphorene along the AC direction ($\varepsilon_{cr}=0.49$, red) and ZZ direction ($\varepsilon_{cr}=0.25$, blue). The dashed lines on the right panel (b) specify the failure stress for pristine phosphorene along the AC direction ($\sigma_{cr}=6.4$ GPa, red) and ZZ direction ($\sigma_{cr}=13.2$ GPa, blue).

As shown in Table 3, the effect of grain boundaries on failure strain and failure stress of phosphorene and graphene are quite similar [52], especially for the samples with the highest GB tilt angles.

Table 3: Comparison of the effect of grain boundaries on failure strain and failure stress of phosphorene and graphene [52].

| Sample orientation | AC | | | | ZZ | | | |
|---|---|---|---|---|---|---|---|---|
| Minimal and maximal reduction of failure strain and stress | $\frac{\Delta\varepsilon}{\varepsilon}$ | | $\frac{\Delta\sigma}{\sigma}$ | | $\frac{\Delta\varepsilon}{\varepsilon}$ | | $\frac{\Delta\sigma}{\sigma}$ | |
| | min | max | min | max | min | max | min | max |
| Phosphorene | -16.3% | -61.2% | -15.1% | -46.9% | -28.0% | -52.0% | -12.9% | -20.9% |
| Graphene | -30.4% | -52.3% | -15.8% | -35.0% | -48.6% | -55.8% | -22.3% | -28.2% |

## 4. Conclusions

Using density functional tight-binding calculations, we studied the structural properties, large deformations and failure mechanism of phosphorene with linear low- and high-angle grain boundaries oriented along the armchair and zigzag directions. We examined the effects of the low and high tilt angles on the failure strain and the failure strength of phosphorene and compared them with other 2D materials. We found that the first critical bonds to rupture at the onset of failure are the most stretched bonds in octagons of (4|8) and heptagons of (5|7) edge dislocations oriented along the tensile direction. These critical bonds are of the same type of bonds in the samples with different tilt angles. They connect the octagons and adjacent hexagons in (4|8) defects, and the heptagons and the adjacent hexagons in (5|7) defects, just as in graphene and hexagonal born-nitride. Phosphorene with high-angle grain boundaries formed by densely packed edge dislocations is much stronger than that with low-angle grain boundaries. Similarly to graphene and hexagonal born-nitride, the higher defect concentration at the grain boundaries does not inevitably lead to greater deterioration of its mechanical properties. The high tilt-angle boundaries are better able to accommodate the pre-strain and prevent failure that originates from the rupture of the critical bonds of the (4|8) and (5|7) edge dislocations. The closer the (4|8) or (5/7) dislocations to each other, the more significant the mutual cancellation of their overlapping compressive and tensile stress fields. Therefore, the tensile pre-strain of the critical bonds is reduced in high-angle grain boundaries. Thus, the initial pre-strain of these critical bonds defines the intrinsic strength. The findings here may be used to modify the mechanical properties of phosphorene via defect engineering, such as intentional introduction and manipulation of grain boundaries, for applications in nanoelectromechanical systems (NEMS).

## 5. Acknowledgements

This work was supported by the A*STAR Computational Resource Centre through the use of its high performance computing facilities, and in part by a grant from the Science and Engineering Research Council (152-70-00017).